\documentclass[useAMS,usenatbib]{mn2e}
\usepackage{graphicx}

\title[Exact polytropic equilibria of self-gravitating systems]
{On exact polytropic equilibria of self-gravitating
gaseous and radiative systems:
their application to molecular cloud condensation}
\author[M. Honda and Y. S. Honda]{Mitsuru Honda$^{1}$
and Yasuko S. Honda$^{2}$\thanks{E-mail: yasuko@ktc.ac.jp}\\
$^{1}$Plasma Astrophysics Laboratory, Institute for Global Science,
Mie 519-5203, Japan\\
$^{2}$Department of Electrical and Information Engineering,
Kinki University Technical College,
Mie 519-4395, Japan}
\begin{document}

\date{Accepted 2003 January 3. Received 2002 November 25;
in original form 2002 November 24}

\pagerange{\pageref{firstpage}--\pageref{lastpage}} \pubyear{2003}

\maketitle

\label{firstpage}

\begin{abstract}
We propose a novel mathematical method to construct
an exact polytropic sphere in self-gravitating
hydrostatic equilibrium,
improving the non-linear Poisson equation.
The central boundary condition for the present equation
requires a ratio of gas pressure to total one at the centre,
which is uniquely identified
by the whole mass and molecular weight of the system.
The special solution derived from the Lane-Emden equation
can be reproduced.
This scheme is now available for modelling
the molecular cloud cores in interstellar media.
The mass-radius relation of the first core is found to be consistent with
the recent results of radiation hydrodynamic simulations.
\end{abstract}


\section{Introduction}

Fundamental research on the structure and evolution of stars,
molecular clouds, and galaxies is still very important \citep{longair94}.
As a rule, a family of the Lane-Emden (LE) equations
classified by the polytrope indices $n$
is known to be the most powerful tool to investigate
the self-gravitating polytropic equilibria \citep{binney87}.
So far, many numerical and semi-analytical works have been
devoted to seeking solutions
\citep{liu96, roxburgh99, goenner00, hunter01, medvedev01},
which include the non-singular function with $n \sim \infty$
for an isothermal polytropic sphere
\citep{natarajan97}.
Removing infinite divergence of the enclosed mass,
some modified models are used to depict the globular clusters,
intergalactic media, clusters of galaxies, and so forth
\citep{king62, jones84}.

Prior to modern numerical methods
\citep{vandenberg85a, vandenberg85b},
the pioneering stellar model proposed by Eddington, which
is based on the solution of the LE equation with the polytrope index of
$n=3$, could be useful to envisage stellar interiors,
although the complexities such as transport processes and
non-ideal equation of state (EOS) have been omitted.
Regarding the study of the interstellar medium (ISM),
it seems that some classes of the LE equations
shed light on dark molecular clouds and their condensations
\citep{mizuno94}.
However, in the general case that
the pressure which supports the celestial objects is
provided largely by the thermal motions of particles,
we need to pay attention to the extensive use of
the LE equation with a specified index.
For heuristic ways, we explain this crucial point below.

Let us suppose that the object
sustains its own self-gravity due to the total pressure of
$P=p_{\rmn{g}}+p_{\rmn{r}}$,
where the partial pressures of gas and radiation field are described by
the EOS of
$p_{\rmn{g}}=\beta P=[ k_{\rmn{B}}/(\mu m_{\rmn{p}}) ] (T/V)$ and
$p_{\rmn{r}}= (1-\beta)P={1 \over 3}aT^4$, respectively.
Here, $m_p$ is the proton mass, $V=\rho^{-1}$ is the specific volume,
$a=8\pi^5 k_{\rmn{B}}^4/(15 c^3 h^3)$ is the Stefan-Boltzmann constant,
and $\mu \gid 1$ denotes the mean molecular weight.
\footnote{For plasmas, $1/2 \lid \mu \simeq 2/(1+3X+0.5Y) \lid 2$,
where the mass fractions contained in the unit mass of the medium are
$X~{\rmn{g}}$ hydrogen, $Y~{\rmn{g}}$ helium, and
$Z=1-X-Y~{\rmn{g}}$ heavy elements.}
For $0<\beta <1$, the relation between temperature and density
is given by $T=\{ [ k_{\rmn{B}}/(\mu m_{\rmn{p}}) ] (3/a)
[ (1-\beta)/\beta ] \}^{1/3}V^{-1/3}$.
Hence, we obtain the polytropic relation as follows:
\begin{equation}
P(r)=\left[ \left( {k_{\rmn{B}}\over {\mu m_{\rmn{p}}}} \right)^4
{3\over a}
{ {1- \beta} \over {\beta^4} }
\right]^{1/3}
V^{-4/3}(r).
\end{equation}
If we assume $\beta = {\rm const}$ \citep{chandrasekhar67},
equation (1) has the form of $P=K\rho^{4/3}$,
where $K=\{ [ k_{\rmn{B}}/(\mu m_{\rmn{p}}) ]^4 ( 3/a )
[ (1- \beta)/\beta^4 ] \}^{1/3}$ is the adiabatic constant.
In this aspect, such a {\it radiative} polytrope
leads to the LE equation of index $n=3$ for gravitational equilibria.
Hereafter, we refer to this scheme, namely,
the Eddington model, as 'LE$3$', without further notice.
Apparently, for {\it gaseous} objects of $\beta \simeq 1$,
equation (1) does not match
with the relation of $P(r) \propto \rho^{5/3}(r)$
involving $n=3/2$ for a perfect monatomic gas.

The possible scenario that recovers the physical consistency is to
allow the spatial variation of polytrope index.
In equation (1) such changes can be achieved by
invoking the density dependence of $\beta = \beta[V(r)]$, and
a trivial solution of ${\rmn{d}} \mu/\mu=0$ (not shown).
Of course, these are not mathematical artefacts.
In the astronomical context,
the present scheme is applicable to study, especially,
the high-density adiabatic cores deeply immersed in
the isothermal molecular cloud,
which are so-called 'first cores' \citep{masunaga98}.
As yet, there has been no distinct detection of the first cores.
We hope for future-planed observations of molecular cloud cores
in the range of millimetre and submillimetre wavelengths
with high angular resolution \citep{hasegawa01}.

In this paper,
we present a self-consistent solution of a polytropic sphere
for self-gravitating gaseous and radiative systems,
going beyond the LE framework.
In the following, we derive a generic equation for hydrostatic equilibria.
In the present scheme,
if we know the whole mass and the mean molecular weight of the system,
then the pressure ratio of gas/total pressure at the centre,
which contributes to the boundary condition, is identified.
It is shown that the polytrope varies radially, modifying
the density and pressure profiles derived from the LE$3$ model.
In the limit that
the gaseous pressure is asymptotically close to the total one,
the solution exactly agrees with the LE solution of index $n=3/2$.

\section[]{The nonlinear Poisson equation
\\* including a self-consistent variable
\\* polytrope}

We begin by defining the adiabatic exponent as
$\Gamma_1={\rmn{d}}({\rmn{ln}} P)/{\rmn{d}}({\rmn{ln}} \rho)$.
Introducing the total derivative of
${\rmn{d}}P={\rmn{d}}(p_{\rmn{g}}+p_{\rmn{r}})
=-p_{\rmn{g}}({\rmn{d}}V/V)+(p_{\rmn{g}} + 4 p_{\rmn{r}})({\rmn{d}}T/T)$,
we obtain
\begin{equation}
(p_{\rmn{g}} + 4 p_{\rmn{r}}){{{\rmn{d}}T} \over T}
-[p_{\rmn{g}} - (p_{\rmn{g}}+p_{\rmn{r}})\Gamma_1] {{{\rmn{d}}V} \over V}=0.
\end{equation}
For a quasi-static and adiabatic change, the thermodynamical principle requires
\begin{equation}
{{\rmn{d}}Q \over V} =
\left({1 \over {\gamma - 1}}p_{\rmn{g}} + 12p_{\rmn{r}} \right)
{ {\rmn{d}}T \over T}
+ (p_{\rmn{g}} + 4p_{\rmn{r}}){ {\rmn{d}} V \over V} \simeq 0,
\end{equation}
where $\gamma=c_{\rmn{p}}/c_{\rmn{v}}$ is the specific-heat ratio.
Combining equation (2) with equation (3),
we find the relation between $\Gamma_1$ and $\beta$ in the form of
\begin{equation}
\Gamma_1[\beta(V), \gamma]=
\beta(V) +
{ {\left( \gamma -1 \right) \left[ {4-3 \beta (V)} \right]^2} \over
{ \beta (V) +12 \left(\gamma- 1\right) \left[1-\beta (V)\right] } }.
\end{equation}
It should be noticed that equation (4) holds the asymptotic properties of
$\Gamma_1 \rightarrow 4/3$
and $\gamma$ for $\beta \rightarrow 0$ and $1$,
respectively.

On the other hand, taking the derivative of equation (1), we obtain
${\rmn{d}}P/{\rmn{d}}V =
-(P/3)\{ ( 4-3\beta )/ [ \beta (1-\beta) ] \}
( {\rmn{d}}\beta/{\rmn{d}}V )
-(4/3)(P/V)$.
This may be rewritten as
\begin{equation}
{{\rm d}P \over P}+{4 \over 3}{ {\rm d}V \over V } +
{ {4-3\beta} \over {3 \left(1-\beta \right)} }
{ {\rm d}\beta \over \beta } =0.
\end{equation}
According to the relation of ${\rm d}P/P + \Gamma_1{\rm d}V/V = 0$,
therefore, we obtain the new relation of
\begin{equation}
V{ {{\rm d}\beta} \over {{\rm d}V} } =
{ {3\beta (1-\beta)} \over {4-3\beta} }
\left[\Gamma_1(\beta, \gamma) - {4\over3} \right],
\end{equation}
which effectively changes the polytrope index of equation (1).

Now we consider a hydrostatic equilibrium of a self-gravitating system,
described by the spherically symmetric Poisson equation, namely,
$\nabla \Phi = r^{-2}({\rm d}/{\rm d}r) (r^2 \Phi)= 4\pi G \rho$,
where $\Phi(r) = -\rho^{-1}({\rm d}P/{\rm d}r)$ and
$G$ is the gravitational constant.
By using equations (1) and (6), the non-linear equation can be cast to

\[
{ 1\over {4\pi G} }
\left[ \left( {k_{\rmn{B}}\over {\mu m_{\rmn{p}}}} \right)^4
{3\over a} \right]^{1/3}
\]

\begin{equation}
\times
{ 1\over {r^2} } { {\rmn{d}}\over {{\rmn{d}}r} }
\left( r^2 \left[ {{1-\beta(\rho)} \over {\beta^4(\rho)}} \right]^{1/3}
{ {\Gamma_1 \left[ \beta (\rho),\gamma \right] } \over \rho^{2/3}(r) }
{ {{\rmn{d}}\rho(r)} \over {{\rmn{d}}r} } \right) = \rho(r).
\end{equation}
In order to normalize equation (7),
we introduce $\tilde{\rho}(\xi) = \rho(\xi)/\rho_{\rmn{c}}$,
where $\rho_{\rmn{c}}$ is the central mass density, and we
properly choose the dimensionless frame of $\xi=r/\alpha$.
Here, the characteristic scalelength $\alpha$ is given by
\begin{equation}
\alpha (\rho_{\rmn{c}},\mu) =
\left[ { 1\over {(4\pi G)^3 \rho_{\rmn{c}}^2} }
\left( {k_{\rmn{B}}\over {\mu m_{\rmn{p}}}} \right)^4
{3\over a} \right]^{1/6}.
\end{equation}
Then, we newly find a set of ordinary differential equations
in the following form:
\[
{1\over{\xi^2}} {{{\rmn{d}}\tilde{m}(\xi)} \over{{\rmn{d}} \xi} }
=-\tilde{\rho}(\xi),
\]

\[
\tilde{m}(\xi)={ \xi^2 \over {\tilde{\rho}^{2/3}(\xi)} }
{ {\rmn{d}}\tilde{\rho}(\xi) \over {{\rmn{d}} \xi} }
\left[
{ {1-\beta (\xi)} \over {\beta^4(\xi)} } \right]^{1/3}
\Gamma_1[\beta(\xi), \gamma],
\]

\begin{equation}
-{1\over {\tilde{\rho}(\xi)} }
{ {\rmn{d}}\tilde{\rho}(\xi) \over {{\rmn{d}} \xi} }
= { {4-3 \beta (\xi)} \over {3\beta (\xi) [1-\beta (\xi)]
\{\Gamma_1[\beta(\xi), \gamma] - {4\over 3}\} } }
{{{\rmn{d}}\beta(\xi)} \over{{\rmn{d}} \xi} }.
\end{equation}
Note that the enclosed mass defined by
$m(r) = 4\pi \int_{0}^{r}\rho(r')r'^2 {\rmn{d}}r'$ is transformed into
$m(\xi) = -4\pi \rho_{\rmn{c}}
\alpha^3(\rho_{\rmn{c}},\mu) \tilde {m}(\xi) \gid 0$.
In the special case of ${\rmn{d}}\beta/{\rmn{d}}\tilde{\rho} \simeq 0$,
equation (9) reduces to the LE equation with an index of
$n \simeq (\gamma-1)^{-1}$ and $3$ for
$\beta \simeq 1$ and $0$, respectively.

\section[]{Numerical solutions of the generalized equation}

To survey the inner structure of the polytropic sphere,
we numerically integrate equation (9) for $\gamma=5/3$, and
the boundary conditions of $\tilde {\rho}(\xi)|_{\xi \rightarrow 0} = 1$,
$\tilde {m}(\xi)|_{\xi \rightarrow 0} = 0$,
and $\beta(\xi)|_{\xi \rightarrow 0} = \beta_{\rmn{c}}$
where $\beta_{\rmn{c}}=(p_{\rmn{g}}/P)_{\rmn{c}}$
is a parameter given at the centre.
It is noted that the solutions of $\tilde {\rho}(\xi)$
and $\tilde {m}(\xi)$ tend to monotonically decrease outwards,
while $\beta(\xi)$ monotonically increases.
At $\xi = \xi_1$, $\tilde {\rho}(\xi)$
and ${\rmn{d}}\tilde {m}(\xi)/{\rmn{d}}\xi$ vanish,
indicating a well-defined radius and mass of the object, i.e.
\begin{equation}
R=\alpha(\rho_{\rmn{c}},\mu)\xi_1(\beta_{\rmn{c}}),
\end{equation}
\begin{equation}
M=m[\xi_1(\beta_{\rmn{c}})]=-4\pi \rho_{\rmn{c}} \alpha^3(\rho_{\rmn{c}},\mu)
\tilde {m}[\xi_1(\beta_{\rmn{c}})].
\end{equation}

\noindent
Table 1 lists the typical values of $\xi_1$ and $\tilde {m}(\xi_1)$
as a function of $\beta_{\rmn{c}}$.

\begin{table}
\centering
 \begin{minipage}{84mm}
  \caption{Values of $\xi_1$, $\tilde {m}$, $\mu$, and $\rho_{\rmn{c}}$
           as a function of $\beta_{\rmn{c}}$.}
  \begin{tabular}{@{}lllll@{}}
  \hline
  $(1-\beta)_{\rmn{c}}$ & $\xi_1$ & $\tilde{m}(\xi_1)$ &
  $\mu \surd{({M \over \sun}) }$
   \footnote{For $\mu=2$, see also Fig.~\ref{f3} (dotted curve).} &
  $\rho_{\rmn{c}} ({\sun \over M})({R \over 100}~\rmn{au})^{3}$ \\
    &   &   &   &
    $\times 10^{-13}~\rmn{g~cm^{-3}}$ \\
 \hline
 $0.01$ & $2.72$ & $-1.10$ & $1.11$ & $8.62$ \\
 $0.05$ & $3.77$ & $-2.76$ & $1.75$ & $9.17$ \\
 $0.1$ & $4.55$ & $-4.49$ & $2.24$ & $9.95$ \\
 $0.3$ & $7.57$ & $-14.2$ & $3.98$ & $14.4$ \\
 $0.5$ & $12.2$ & $-38.6$ & $6.57$ & $22.1$ \\
 $0.7$ & $21.6$ & $-1.36\times 10^2$ & $12.3$ & $35.3$ \\
 $0.9$ & $56.8$ & $-1.48\times 10^3$ & $40.7$ & $58.6$ \\
\hline
\end{tabular}
\end{minipage}
\end{table}

\subsection{Inner structure of gaseous and radiative
polytropic sphere}

Substituting equation (8) into equation (11),
we obtain the explicit notation of the mean molecular weight:
$\mu = [(4\pi G^3 M^2)^{-1}(k_{\rmn{B}}/m_{\rmn{p}})^4(3/a)]^{1/4}
\{-\tilde{m}[\xi_1(\beta_{\rmn{c}})]\}^{1/2}$.
Taking the allowed parameter region into consideration,
this can be expressed as
\begin{equation}
\mu(M,\beta_{\rmn{c}}) = 1.06
\left( M_{\sun} \over M \right)^{1/2}
\{-\tilde{m}[\xi_1(\beta_{\rmn{c}})]\}^{1/2} \gid 1,
\end{equation}
which corresponds to the LE3 solution of
$\mu_{\rmn{LE3}} = 4.25 (M_{\sun}/M)^{1/2}
(1-\beta_{\rmn{c}})^{1/4}/\beta_{\rmn{c}}$.
Furthermore, using equations (8) and (10), we obtain
$\rho_{\rmn{c}} = \{(4\pi G R^2)^{-3}
[k_{\rmn{B}}/(\mu m_{\rmn{p}})]^4(3/a)\}^{1/2}
\xi_1^3(\beta_{\rmn{c}})$.
Making use of expression (12), this can be written as
\noindent
\[
\rho_{\rmn{c}}(M,R,\beta_{\rmn{c}}) = 4.73 \times 10^{-14}
\]
\begin{equation}
\times 
\left( {M \over M_{\sun}} \right)
\left(100~\rmn{au} \over R \right)^3
{ {\xi_1^3(\beta_{\rmn{c}})} \over
{ \{ -\tilde{m}[\xi_1(\beta_{\rmn{c}})] \}} }
~{\rmn{g~cm^{-3}}}.
\end{equation}
Thus, for given $M$ and $\mu$, the fraction of gas pressure
$\beta_{\rmn{c}}$ is identified by equation (12).
In addition, for given $R$, the central mass density is determined by
equation (13), in contrast to the LE3 solution of
$\rho_{\rmn{c~LE3}} =
7.68 \times 10^{-12} (M/M_{\sun}) (100~\rmn{au}/R)^3~{\rmn{g~cm^{-3}}}$,
which is independent on $\beta_{\rmn{c}}$.
These dependencies are summarized in Table~1.

\begin{figure}
\includegraphics[width=108mm]{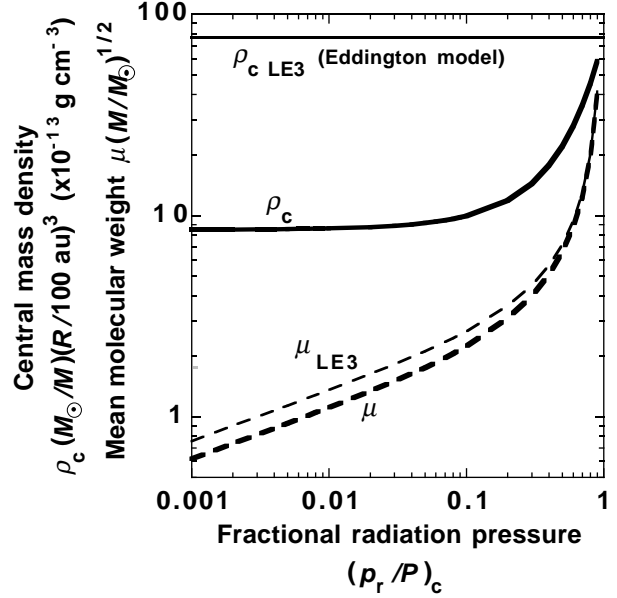}
\caption{Central mass density of
$\rho_{\rmn{c}}$ (bold solid curve) and
$\rho_{\rmn{c~LE3}}$ (solid line),
both multiplied by $(M_{\sun}/M) (R/100~\rmn{au})^3$,
and mean molecular weight of $\mu$ (bold dashed curve)
and $\mu_{\rmn{LE3}}$ (dashed curve),
both multiplied by $(M/M_{\sun})^{1/2}$,
as a function of the central pressure ratio of
$(p_{\rmn{r}}/P)_{\rmn{c}}$. For an explanation, see the text.}
\label{f1}
\end{figure}

In Fig.~\ref{f1}, we plot equations (12) and (13) as a function of
$(p_{\rmn{r}}/P)_{\rmn{c}}=(1-\beta)_{\rmn{c}}$.
It is found that, for $M=M_{\sun}$ as an example,
the inequality of equation (12) requires
the fraction of radiation pressure at the centre to take
its parameter region of $(1-\beta)_{\rmn{c}} \gid 6.7 \times 10^{-3}$.
\footnote{For plasmas, replace with
$4.4 \times 10^{-4} \lid (1-\beta)_{\rmn{c}} \lse 7.4 \times 10^{-2}$,
corresponding to the allowed parameter region of $1/2 \lid \mu \lse 2$
(see footnote~1).}
If we specify the mean molecular weight,
then the fraction can be fixed numerically.
For $\mu = 2$ (hydrogen molecule), we obtain
$(1-\beta)_{\rmn{c}} = 7.4 \times 10^{-2}$.
In such a gas dominant regime of $(1-\beta)_{\rmn{c}} \ll 1$,
the dimensionless quantity $\sim \rho_{\rmn{c}}(R^3/M)$
does not largely depend on $\beta_{\rmn{c}}$, as seen in the figure.
For $R=100~\rmn{au}$, the numerical solution reveals the central density of
$\rho_{\rmn{c}} = 8.6  \times 10^{-13}~{\rmn{g~cm^{-3}}}$,
which is about $90$ per cent lower than that of
$\rho_{\rmn{c~LE3}} = 7.68 \times 10^{-12}~{\rmn{g~cm^{-3}}}$.
For massive objects of $M \gg M_{\sun}$, the radiation dominant regime appears.
In the limit of $\beta_c \rightarrow 0$,
both $\rho_{\rmn{c}}$ and $\mu$ are asymptotically close to
$\rho_{\rmn{c~LE3}}$ and $\mu_{\rmn{LE3}}$, respectively.
This characteristic is a result of the behavior of
$\Gamma_1 \rightarrow 4/3$ for $\beta \rightarrow 0$ in equation (4).

\begin{figure}
\includegraphics[width=90mm]{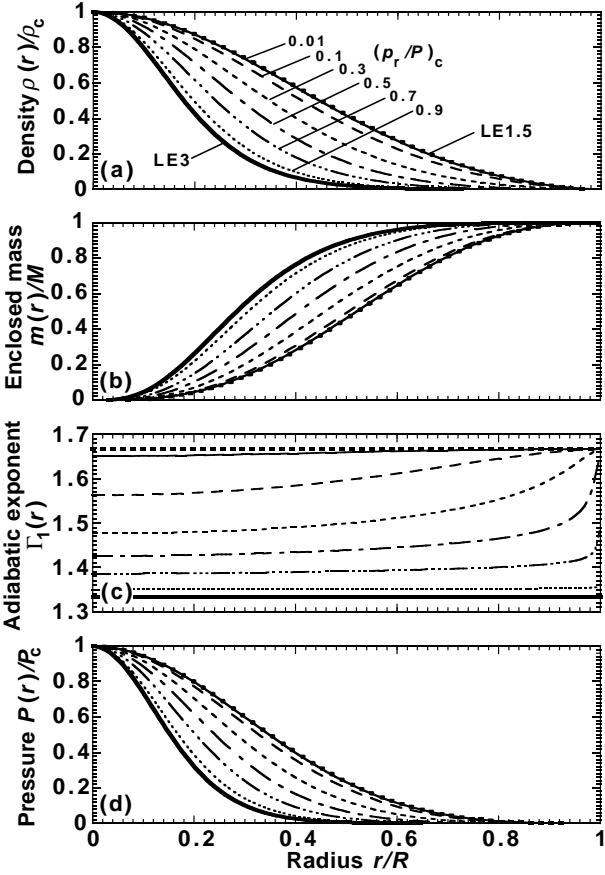}
\caption{Radial profiles of
(a) normalized mass density $\rho (r)/\rho_{\rmn{c}}$,
(b) normalized enclosed mass $m(r)/M$,
(c) adiabatic exponent $\Gamma_1(r)$, and
(d) normalized total pressure $P(r)/P_{\rmn{c}}$,
for $(p_r/P)_{\rmn{c}} = 0.01$ (solid curves),
$0.1$ (long-dashed curves), $0.3$ (short-dashed curves),
$0.5$ (dot-dashed curves), $0.7$ (triple dot-dashed curves)
and $0.9$ (dotted curves).
The figures have a common horizontal axis indicating
the normalized radius of $r/R$.
For comparison, the LE solutions with the polytrope index of
$n=3$ (LE3; bold solid curves/line) and
$n=3/2$ (LE1.5; bold dotted curves/line) are also plotted.
Note that in (a), (b) and (d) the solid curves for $(p_r/P)_{\rmn{c}} = 0.01$
overlap the bold dotted curves for the LE1.5 solution (see text).}
\label{f2}
\end{figure}

In Fig.~\ref{f2}, for $0.01 \lid (1-\beta)_{\rmn{c}} \lid 0.9$,
we show the radial profiles of the
normalized mass density $\rho(r)/\rho_{\rmn{c}}$,
enclosed mass $m(r)/M$, total pressure $P(r)/P_{\rmn{c}}$, as well as
the adiabatic exponent $\Gamma_1(r)$,
comparing the LE solutions of index $n=3$ and $3/2$.
Here the radius $r=\alpha \xi$ is normalized by $R$.
It is shown in Fig.~\ref{f2}(a)
that the density profile gradually deviates from
that of the LE3 solution as $(1-\beta)_{\rmn{c}}$ decreases.
Concerning the case that the parameters of $M$ and $R$ are fixed,
in the lower $(1-\beta)_{\rmn{c}}$ the central density decreases,
because the density varies sufficiently slowly from the centre to the envelope.
In the limit of $\beta_{\rmn{c}} \rightarrow 1$,
the present solution shows an exact agreement with
that of the LE equation with a polytrope index of $n=3/2$.
The radial profiles of the enclosed mass are shown in Fig.~\ref{f2}(b).
For higher $(1-\beta)_{\rmn{c}}$ including the LE3,
the mass is more condensed in the central region.
For example, the radius confining the half-mass
indicates $R_{\rmn{M/2~LE3}} = 0.325R$.
As $(1-\beta)_c$ decreases, this radius gradually shifts outward,
up to $R_{\rmn{M/2~LE1.5}} = 0.521R$.
It is confirmed that the gaseous polytropic effects
are likely to flatten the density profile significantly.

In Fig.~\ref{f2}(c), we show the radial profiles of
$\Gamma_1[\beta(\xi),\gamma=5/3]$, defined by equation (4).
As the argument $\beta(\xi)$ monotonically increases outwards,
the adiabatic exponent also monotonically increases,
yielding the self-consistent variable polytrope.
This is one of the most important results in the present paper.
For all $\beta_{\rmn{c}}$ being considered,
the exponent varies quite slowly near the central region,
while for $(1-\beta)_{\rmn{c}} \goa 0.5$,
the variation around the envelope tends to be very steep.

In Fig.~\ref{f2}(d), we show the radial profiles of total pressure of
$P(\xi) = [G M^2/(4 \pi R^4)]
\{ -\xi_1^2(\beta_{\rmn{c}})/\tilde{m}[\xi_1(\beta_{\rmn{c}})]\}^2
\{[1-\beta(\xi)]/\beta^4(\xi) \}^{1/3} \tilde{\rho}^{4/3}(\xi)$,
normalized by the central pressure $P_c$.
It is noteworthy that, for the {\it gas} dominant regime of
$(1-\beta)_{\rmn{c}} \ll 1$,
the pressure formula exhibits the density dependence approximated by
$P(\xi) \propto \tilde{\rho}^{5/3}(\xi)$
with $\Gamma_1 \simeq \gamma$,
fully consistent with Fig.~\ref{f2}(c).
At $\xi=0$, the pressure takes the peak value, to give
\noindent
\[
P_{\rmn{c}} = 4.19 \times 10^{-3}
\left(M \over M_{\sun} \right)^2
\left(100~\rmn{au} \over R \right)^4
\]
\begin{equation}
\times
{ {\xi_1^4(\beta_{\rmn{c}})} \over {\tilde{m}^2[\xi_1(\beta_{\rmn{c}})]} }
\left[ { (1-\beta_{\rmn{c}}) \over {\beta_{\rmn{c}}^4} } \right]^{1/3}
~{\rmn{ergs~cm^{-3}}}.
\end{equation}
Notice that, in contrast with the LE3 solution of
$P_{\rmn{c~LE3}} = 0.582
(M/M_{\sun})^2 (100~\rmn{au}/R)^4~{\rmn{ergs~cm^{-3}}}$,
equation (14) does depend upon $\beta_{\rmn{c}}$,
but weakly for $(1-\beta)_{\rmn{c}} \ll 1$.
For $M=M_{\sun}$, $\mu = 2$, and $R=100~\rmn{au}$,
we obtain the central pressure of
$P_{\rmn{c}} = 4.6 \times 10^{-2}~{\rmn{ergs~cm^{-3}}}$,
which is an order of magnitude smaller than $P_{\rmn{c~LE3}}$.

Invoking the EOS, the temperature profile can be described as
$T(\xi) = [3 G M^2/(4 \pi a R^4)]^{1/4}\xi_1(\beta_{\rmn{c}})$
$\times
\{ -\tilde{m}[\xi_1(\beta_{\rmn{c}})] \}^{-1/2}
\{[1-\beta(\xi)]/\beta(\xi) \}^{1/3} \tilde{\rho}^{1/3}(\xi)$
(not shown in figure).
At $\xi=0$, it takes the peak value of
\[
T_{\rmn{c}} = 1.14 \times 10^{3}
\left(M \over M_{\sun} \right)^{1/2}
\left( {100~\rmn{au} \over R} \right)
\]
\begin{equation}
\times
{ {\xi_1(\beta_{\rmn{c}})} \over
\{ -\tilde{m}[\xi_1(\beta_{\rmn{c}})] \}^{1/2} }
\left[ { (1-\beta_{\rmn{c}}) \over {\beta_{\rmn{c}}} } \right]^{1/3}
~{\rmn{K}},
\end{equation}
whereas the LE3 solution reads $T_{\rmn{c~LE3}} = 3.90 \times 10^{3}
(M/M_{\sun})^{1/2} (100~\rmn{au}/R) (1-\beta_{\rmn{c}})^{1/4}~{\rmn{K}}$;
both having the dependence of $\beta_{\rmn{c}}$.
For $M=M_{\sun}$, $\mu = 2$, and $R=100~\rmn{au}$,
the central temperature of equation (15)
is found to be $T_{\rmn{c}} = 1.1 \times 10^{3}~{\rmn{K}}$,
which is again lower than
$T_{\rmn{c~LE3}} = 2.0 \times 10^{3}~{\rmn{K}}$.
It is found that the decrease of the temperature is
relatively small, within a factor of 2 for each $\beta_{\rmn{c}}$.

\subsection{Their application to the molecular cloud condensation:
the critical radius of the 'first core'}

In the context of the study of molecular cloud condensation in the ISM,
the present quasi-stationary model is now available
to provide an insight into the complicated dynamics
\citep{penston69, larson69, shu77, saigo98},
in particular, the formation of the first core (Masunaga et al. 1998).
For such an application, the effects of magnetic fields and turbulence
might be taken into account, and the additional pressure $\delta P$
might be effectively included in equation (1)
by replacing $\mu$ with $\mu^{\prime}=\mu \beta_\rmn{p}/(1+\beta_\rmn{p})$,
where $\beta_\rmn{p} = p_\rmn{g}/\delta P > 0$ \citep{bludman96}; e.g.
for pure magnetic pressure, $\beta_\rmn{p} = (8\pi p_\rmn{g})/B^2$.
Moreover, for $T_{\rmn{c}} \goa 2000~\rmn{K}$,
the dissociation of hydrogen molecules triggers
gravitational contraction of the quasi-stationary first core.
By invoking the scaling of $T_\rmn{c} \propto 1/\mu^{\prime}$
and equation (15),
therefore, we find the relation between the mass and radius of
the first core as follows:

\[
R \goa 13
\left( 1+{\beta_\rmn{p} } \over \beta_\rmn{p} \right)
\left(M \over {0.05 M_{\sun}} \right)^{1/2}
\]
\begin{equation}
\times
{ {\xi_1(\beta_{\rmn{c}})} \over
\{ -\tilde{m}[\xi_1(\beta_{\rmn{c}})] \}^{1/2} }
\left[ { (1-\beta_{\rmn{c}}) \over {\beta_{\rmn{c}}} } \right]^{1/3}
~{\rmn{au}},
\end{equation}

\noindent
for $\mu=2$.
Note that the constraint of equation (12) gives the dependence of
$\beta_\rmn{c}=\beta_\rmn{c}(M,\mu=2)$ in equation (16).
On the other hand, in the LE3 we obtain the mass-radius
({\it M$-$R}) relation of
$R \goa 4.6 (M/{0.05 M_{\sun}}) \beta_\rmn{c}~\rmn{au}$,
where $\beta_\rmn{c} = \beta_\rmn{c}(M,\mu_\rmn{LE3}=2)$.

\begin{figure}
\includegraphics[width=105mm]{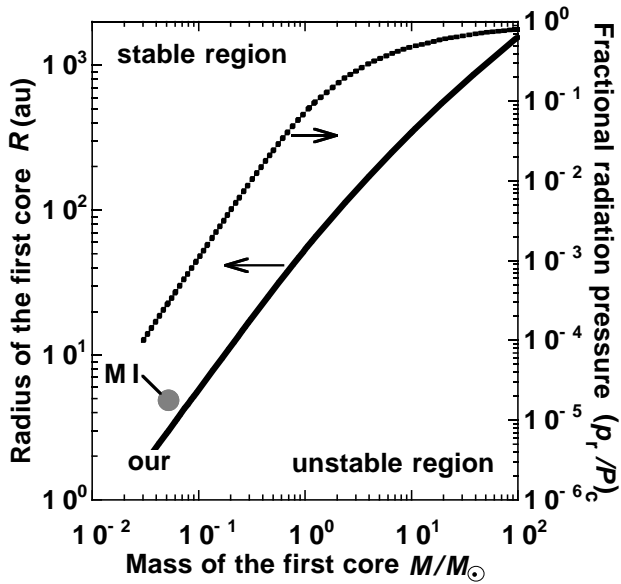}
\caption{Minimum possible radius of the first core of
hydrogen molecular clouds (solid curve)
and fractional radiation pressure at the centre (dotted curve)
as a function of the mass of the core.
The upper-left side of the solid curve
($R > R_\rmn{min}$) corresponds to the stable region,
while the opposite side ($R < R_\rmn{min}$) corresponds to unstable region,
concomitant with the dissociation of hydrogen molecules.
For comparison, the {\it M$-$R} relation
from radiation hydrodynamic simulations
is indicated by the shaded inset (Masunaga et al. 1998).}
\label{f3}
\end{figure}

In Fig.~\ref{f3} for $\beta_\rmn{p} \gg 1$,
we plot the numerical solution of equation (16), that is,
the allowable parameter region in the {\it M$-$R} plane.
For convenience sake, we also plot the
fractional radiation pressure at the centre
as a function of the mass of the core:  
$(p_{\rmn{r}}/P)_{\rmn{c}} = 1 - \beta_\rmn{c}(M,\mu=2)$.
For $M \gg M_{\sun}$,
corresponding to $\beta_\rmn{c} \ll 1$ for fixed $\mu$ (compare Table~1),
the minimum core radius of equation (16) can be
well approximated by that from the LE3:
$R_\rmn{min} \approx 1.9 \times 10^3 [(1+\beta_\rmn{p})/\beta_\rmn{p}]
(M/{100 M_{\sun}})^{1/2}~\rmn{au}$.
On the other hand, for $M \ll M_{\sun}$, we numerically find that
the minimum possible radius of equation (16) scales as

\begin{equation}
R_\rmn{min} \approx 3.0
\left( 1+{\beta_\rmn{p} } \over \beta_\rmn{p} \right)
\left( M \over {0.05 M_{\sun}} \right)~\rmn{au},
\end{equation}

\noindent
and $(1-\beta)_{\rmn{c}} \approx 2.8 \times 10^{-4}
[\beta_\rmn{p}/(1+\beta_\rmn{p})]^4 (M/{0.05 M_{\sun}})^2$.
The estimation of (17) reasonably supports the results of
radiation hydrodynamic calculations by Masunaga et al. (1998).
They show that the mass and radius of the first core are
$M \sim 0.05 M_{\sun}$ and $R \sim 5~\rmn{au} > R_\rmn{min}$,
respectively,
and the results do not largely depend on the initial conditions of
wide parameter ranges.
If $R \loa R_\rmn{min}$, the core tends to be gravitationally unstable and
evolves along the dynamically-contracting track,
to self-organize the 'second core'
as a protostar \citep{masunaga00}.

\section[]{Concluding remarks}

In conclusion, we have developed a generic scheme to construct
an exact polytropic sphere of self-gravitating
gaseous and radiative medium.
The particular results derived from the
newly modified Poisson equation for hydrostatic equilibria are:

\begin{enumerate}
\item the numerical solutions show that for all cases,
the central density and pressure are
lower than those from the LE function with the index of $n=3$;
\item the adiabatic exponent monotonically increases radially outwards; and
\item in the special case that the central polytrope is gaseous closely,
the present solution reproduces the properties of the
LE function with $n=3/2$.
\end{enumerate}

\noindent
Within this framework, the whole mass of the system is connected with
the central density, temperature, and the mean molecular weight.

For an application to modelling the molecular cloud condensation in the ISM,
we have newly found the scaling law of the critical radius of the first core.
The preliminary result is in consistent with that of
the radiation hydrodynamic simulations.
We expect that the major consequence can be also referred to,
for example, {\it mutatis mutandis},
the study of the passive phase of protoplanetary discs \citep{honda99},
stellar modelling \citep{basu00}, and so on.

\section*{Acknowledgments}

We acknowledge the Information and Educational Center,
Kinki University TC for their hospitality.
YSH is grateful to Yoshitsugu~Nakagawa for a useful discussion.

\bsp

\label{lastpage}


\begin{thebibliography}{99}

\bibitem[\protect\citeauthoryear{Basu, Pinsonneault \& Bahcall}{2000}]{basu00}
Basu~S., Pinsonneault~M.~H., Bahcall~J.~N., 2000, ApJ, 529, 1084

\bibitem[\protect\citeauthoryear{Binney \& Tremaine}{1987}]{binney87}
Binney~J., Tremaine~S., 1987, Galactic Dynamics. Princeton Univ. Press,
Princeton, NJ

\bibitem[\protect\citeauthoryear{Bludman \& Kennedy}{1996}]{bludman96}
Bludman~S.~A., Kennedy~D.~C., 1996, ApJ, 472, 412

\bibitem[\protect\citeauthoryear{Chandrasekhar}{1967}]{chandrasekhar67}
Chandrasekhar~S., 1967, An Introduction to the Study of Stellar Structure.
Dover, New York

\bibitem[\protect\citeauthoryear{Goenner \& Havas}{2000}]{goenner00}
Goenner~H., Havas~P., 2000, J.~Math.~Phys., 41, 7029

\bibitem[\protect\citeauthoryear{Hasegawa}{2001}]{hasegawa01}
Hasegawa~T., 2001, The Astronomical Herald, 94, 586, in Japanese

\bibitem[\protect\citeauthoryear{Honda \& Nakagawa}{1999}]{honda99}
Honda~Y.~S., Nakagawa~Y., 1999, in Nakamoto~T., ed., Proc. Star Formation.
Nobeyama Radio Observatory. Nobeyama, p.235

\bibitem[\protect\citeauthoryear{Hunter}{2001}]{hunter01}
Hunter~C., 2001, MNRAS, 328, 839

\bibitem[\protect\citeauthoryear{Jones \& Forman}{1984}]{jones84}
Jones~C., Forman~W., 1984, ApJ, 276, 38

\bibitem[\protect\citeauthoryear{King}{1962}]{king62}
King~I.~R., 1962, AJ, 67, 471

\bibitem[\protect\citeauthoryear{Larson}{1969}]{larson69}
Larson~R.~B., 1969, MNRAS, 145, 271

\bibitem[\protect\citeauthoryear{Liu}{1996}]{liu96}
Liu~F.~K., 1996, MNRAS, 281, 1197

\bibitem[\protect\citeauthoryear{Longair}{1994}]{longair94}
Longair~M.~S., 1994, High Energy Astrophysics Vol.2,
Stars, the Galaxy and the interstellar medium.
Cambridge Univ. Press, Cambridge

\bibitem[\protect\citeauthoryear{Masunaga \& Inutsuka}{2000}]{masunaga00}
Masunaga~H., Inutsuka~S., 2000, ApJ, 531, 350

\bibitem[\protect\citeauthoryear
{Masunaga, Miyama \& Inutsuka}{1998}]{masunaga98}
Masunaga~H., Miyama~S.~M., Inutsuka~S., 1998, ApJ, 495, 346

\bibitem[\protect\citeauthoryear{Medvedev \& Rybicki}{2001}]{medvedev01}
Medvedev~M.~V., Rybicki~G., 2001, ApJ, 555, 863

\bibitem[\protect\citeauthoryear{Mizuno et al.}{1994}]{mizuno94}
Mizuno~A., Onishi~T., Hayashi~M., Ohashi~N., Sunada~K.,
Hasegawa~T., Fukui~Y., 1994, Nat, 368, 719

\bibitem[\protect\citeauthoryear
{e.g., Natarajan \& Lynden-Bell}{1997}]{natarajan97}
Natarajan~P., Lynden-Bell~D., 1997, MNRAS, 286, 268

\bibitem[\protect\citeauthoryear{Penston}{1969}]{penston69}
Penston~M.~V., 1969, MNRAS, 114, 425

\bibitem[\protect\citeauthoryear{Roxburgh \& Stockman}{1999}]{roxburgh99}
Roxburgh~I.~W., Stockman~L.~M., 1999, MNRAS, 303, 466

\bibitem[\protect\citeauthoryear{Saigo \& Hanawa}{1998}]{saigo98}
Saigo~K., Hanawa~T., 1998, ApJ, 493, 342

\bibitem[\protect\citeauthoryear{Shu}{1977}]{shu77}
Shu~F.~H., 1977, ApJ, 214, 488

\bibitem[\protect\citeauthoryear{VandenBerg}{1985}]{vandenberg85b}
VandenBerg~D.~A., 1985, ApJS, 58, 711

\bibitem[\protect\citeauthoryear{VandenBerg \& Bell}{1985}]{vandenberg85a}
VandenBerg~D.~A., Bell~R.~A., 1985, ApJS, 58, 561

\end{thebibliography}
\end{document}